              \documentclass[letter,traditabstract]{aa} 
\usepackage{graphicx,natbib}
\usepackage{txfonts}
\def\ts     {\thinspace}
\def\msol   {\ifmmode{{\rm M}_{\odot} }\else{M$_{\odot}$}\fi}
\def\lsol   {\ifmmode{L_{\odot}}\else{$L_{\odot}$}\fi}
\def\lfir   {\ifmmode{L_{\rm FIR}}\else{$L_{\rm FIR}$}\fi}
\def\etal   {{\rm et\ts al.}}
\def\microns {\ifmmode{\mu{\rm m}}\else{$\mu$m}\fi}

\begin{document}
   \title{An extended {\it Herschel} drop-out source in the center of AS1063:\\ a 'normal' dusty galaxy at $z=6.1$ or SZ substructures?}
	\titlerunning{Discovery of a strongly lensed {\it Herschel} drop-out}

   \author{F. Boone
          \inst{1, 2}
          \and
          B.\,Cl\'ement
          \inst{3}
          \and
          J.\,Richard
          \inst{4}
          \and 
          D.\,Schaerer
          \inst{5,2}
          \and
          D.\,Lutz
          \inst{6}         
          \and
          A.\,Wei\ss
          \inst{7}
          \and
          M.\,Zemcov
          \inst{8}
          \and
          E.\,Egami
          \inst{3}
          \and
          T.\,D.\,Rawle
          \inst{9}
           \and
          G.\,L.\,Walth
          \inst{3}
          \and
          J.-P.\,Kneib
          \inst{10,11}
          \and
          F.\,Combes
          \inst{12}
          \and
          I.\,Smail
          \inst{13}
          \and
          A. M.\,Swinbank
          \inst{13}
          \and
          B.\,Altieri
          \inst{9}
          \and
          A. W. Blain
          \inst{14}
          \and
          S.\,Chapman
          \inst{15}
          \and
          M.\,Dessauges-Zavadsky
          \inst{5}
          \and
          R.\,J.\,Ivison
          \inst{16}
          \and	
          K. K. Knudsen
          \inst{17}
          \and	
          A.\,Omont
          \inst{18}
          \and
          R.\,Pell\'o
          \inst{1,2}
  		  \and
          P.\,G.\,P{\'e}rez-Gonz{\'a}lez
          \inst{19}
          \and
          I.\,Valtchanov
          \inst{9}
          \and
          P. van der Werf
          \inst{20}
          \and
          M.\,Zamojski
          \inst{5}
          }

   \institute{ Universit\'e de Toulouse; UPS-OMP; IRAP; Toulouse, France
     \and
     CNRS; IRAP; 9 Av. colonel Roche, BP 44346, F-31028 Toulouse cedex 4, France
     \and
     Steward Observatory, University of Arizona, 933 North Cherry Avenue, Tucson, AZ 85721, USA
     \and
     Centre de Recherche Astrophysique de Lyon, Universit\'e Lyon 1, 9 Avenue Charles Andr\'e, F-69561 Saint Genis Laval, France
	 \and
	 Geneva Observatory, Universit\'e de Gen\`eve, 51 chemin des Maillettes, 1290 Versoix, Switzerland
	 \and
	 Max-Planck-Institut f\"ur extraterrestrische Physik,  Postfach 1312, 85741 Garching, Germany 
     \and
     Max-Planck-Institut f\"ur Radioastronomie, Auf dem H\"ugel 69, 53121 Bonn, Germany
     \and
     Department of Physics, Mathematics, and Astronomy, California Institute of Technology, Pasadena, CA 91125, USA
     \and
     European Space Astronomy Centre (ESAC)/ESA, Villanueva de la Ca\~nada, E-28691 Madrid, Spain
     \and
     Laboratoire d'astrophysique, Ecole Polytechnique F\'ed\'erale de Lausanne, Observatoire de Sauverny, 1290 Versoix, Switzerland
     \and
     Aix Marseille Universit\'e, CNRS, LAM, UMR 7326, 13388, Marseille, France
     \and
     LERMA, Observatoire de Paris, 61 avenue de l'Observatoire, 75014 Paris, France
     \and
     Institute for Computational Cosmology, Department of Physics, Durham University, Durham DH1 3LE, UK
     \and
     Physics \& Astronomy, University of Leicester, Leicester, LE1 7RH, UK
     \and
     Department of Physics and Atmospheric Science, Dalhousie University Halifax, NS, B3H 3J5, Canada
     \and
     Institute for Astronomy, University of Edinburgh, Royal Observatory, Blackford Hill, Edinburgh EH9 3HJ, UK
	 \and
	 Department of Earth and Space Science, Chalmers University of Technology, Onsala Space Observatory, 43992 Onsala, Sweden
	 \and
     UPMC Univ Paris 6, UMR 7095, Institut d'Astrophysique de Paris, 75014, Paris, France
     \and
Departamento de Astrof\'{\i}sica, Facultad de CC. F\'{\i}sicas, Universidad Complutense de Madrid, 28040 Madrid, Spain
	\and      
	Leiden Observatory, Leiden University, P.O. box 9513, 2300 RA Leiden, The Netherlands
     }

   \date{;}
 
  \abstract 
   {  In the course of our 870$\mu$m APEX/LABOCA follow up of the {\it Herschel} Lensing Survey we have detected a source in AS1063 (RXC J2248.7-4431), that has no counterparts in any of the {\it Herschel} PACS/SPIRE bands, it is a {\it Herschel} 'drop-out' with $S_{\rm 870}/S_{\rm 500}\ge$0.5. The 870$\mu$m emission is extended and centered on the brightest cluster galaxy suggesting { either a multiply imaged background source or substructure in the Sunyaev-Zel'dovich (SZ)  increment due to inhomogeneities in the hot cluster gas of this merging cluster. We discuss both interpretations with emphasis on the putative lensed source.} 
   Based on the observed properties and on our lens model we find that this source could be the first SMG with a moderate far infrared luminosity ($L_{\rm FIR}<10^{12} L_{\odot}$)  detected so far at $z>4$. In deep {\it HST} observations we identified a multiply imaged $z\sim6$ source and we measured its spectroscopic redshift $z$=6.107 with VLT/FORS. This source could be associated with the putative SMG but it is most likely offset spatially by 10-30\,kpc and they could be interacting galaxies.  With a FIR luminosity in the range $[5-15]\times 10^{11} L_{\odot}$ corresponding to a star formation rate in the range $[80-260]$\,$M_{\odot}$\,yr$^{-1}$, this SMG would be more representative  than the extreme starbursts usually detected at $z>4$. With a total magnification of $\sim$25 it would open a unique window to the 'normal' dusty galaxies at the end of the epoch of reionization. 
   }


   \maketitle
%

\section{Introduction}
Estimating the contribution of dust obscured star formation in the early Universe is essential to constrain the models of galaxy evolution and it has been a growing field of research since the late 1990s \citep[e.g., ][]{2002PhR...369..111B}.
Submillimeter (submm) surveys have proven to be very efficient at detecting  distant dusty galaxies (also known as Submillimeter Galaxies, SMGs)  owing to the negative K-correction and their redshift distribution  was found to peak at $z\sim2-3$ \citep{2005ApJ...622..772C}.  

With the advent of a new generation of submm instruments 
the hunt for the highest redshift SMGs progressed at a rapid pace in recent years.  The first SMG beyond $z=5$ was discovered by \citet{2011Natur.470..233C} with JCMT/AzTEC. Based on a {\it Herschel} detection and a 30m/EMIR follow up,  \citet{2012A&A...538L...4C} discovered an interacting system of bright SMGs at $z=5.243$. At the same time \citet{2012Natur.486..233W} using IRAM instruments found that an SMG known for years in the Hubble deep field is actually a system of galaxies lying at $z=5.2$.  Following up on SPT bolometer observations \citet{2013Natur.495..344V}, and \citet{2013ApJ...767...88W} measured with ALMA  the spectroscopic redshifts of 23 new SMGs out of which two are at $z>5$. In parallel, \citet{2013Natur.496..329R}  observed 'red' SMGs based on {\it Herschel} colors with CARMA and discovered the highest redshift SMG at $z=6.34$.

In terms of luminosity, however, all the SMGs detected so far beyond $z>4$ are ultraluminous infrared galaxies\footnote{The lowest luminosity SMG detected so far at $z>$4 has $L_{\rm FIR}=1.3\times 10^{12}$\,$L_{\odot}$ and is at $z=4.04$ \citep{2010ApJ...709..210K}} (ULIRGs) with $L_{\rm FIR}>10^{12}$\,$L_{\odot}$ or even hyperluminous infrared galaxies (HyLIRGs) with $L_{\rm FIR}>10^{13}$\,$L_{\odot}$, implying star formation rates $\gtrsim 10^3$\,$M_{\odot}$\,yr$^{-1}$. As confirmed by recent ALMA number counts \citep{2013MNRAS.432....2K, 2013ApJ...769L..27H} these extreme starbursts are not representative of the average population of dusty star forming galaxies at $z>4$ and the luminous infrared galaxies (LIRGs) with $L_{\rm FIR}\sim 10^{11}$\,$L_{\odot}$  that should represent the majority remain to be discovered. {The lensing power provided by massive galaxy clusters is widely used to detect distant galaxies \citep[e.g., ][]{1997ApJ...490L...5S}. However, the recent discovery of substructures in the Sunyaev-Zel'dovich (SZ) increment of interacting clusters \citep{2011ApJ...734...10K, 2012ApJ...761...47M} may complicate the interpretation of submm observations.}
We report here the discovery of a good candidate for a  normal star forming galaxy at $z=6.1$ lensed by the cluster AS1063 (RXCJ2248.7-4431) and we discuss the possibility that this source could instead correspond to substructures in the SZ effect. We adopt the $\Lambda$CDM concordance cosmology: $H_0=71$\,km\,s$^{-1}$\,Mpc$^{-1}$, $\Omega_M=0.27$ and $\Omega_{\Lambda}=0.73$.


\begin{figure*}
\centerline{
\includegraphics[width=16cm,clip]{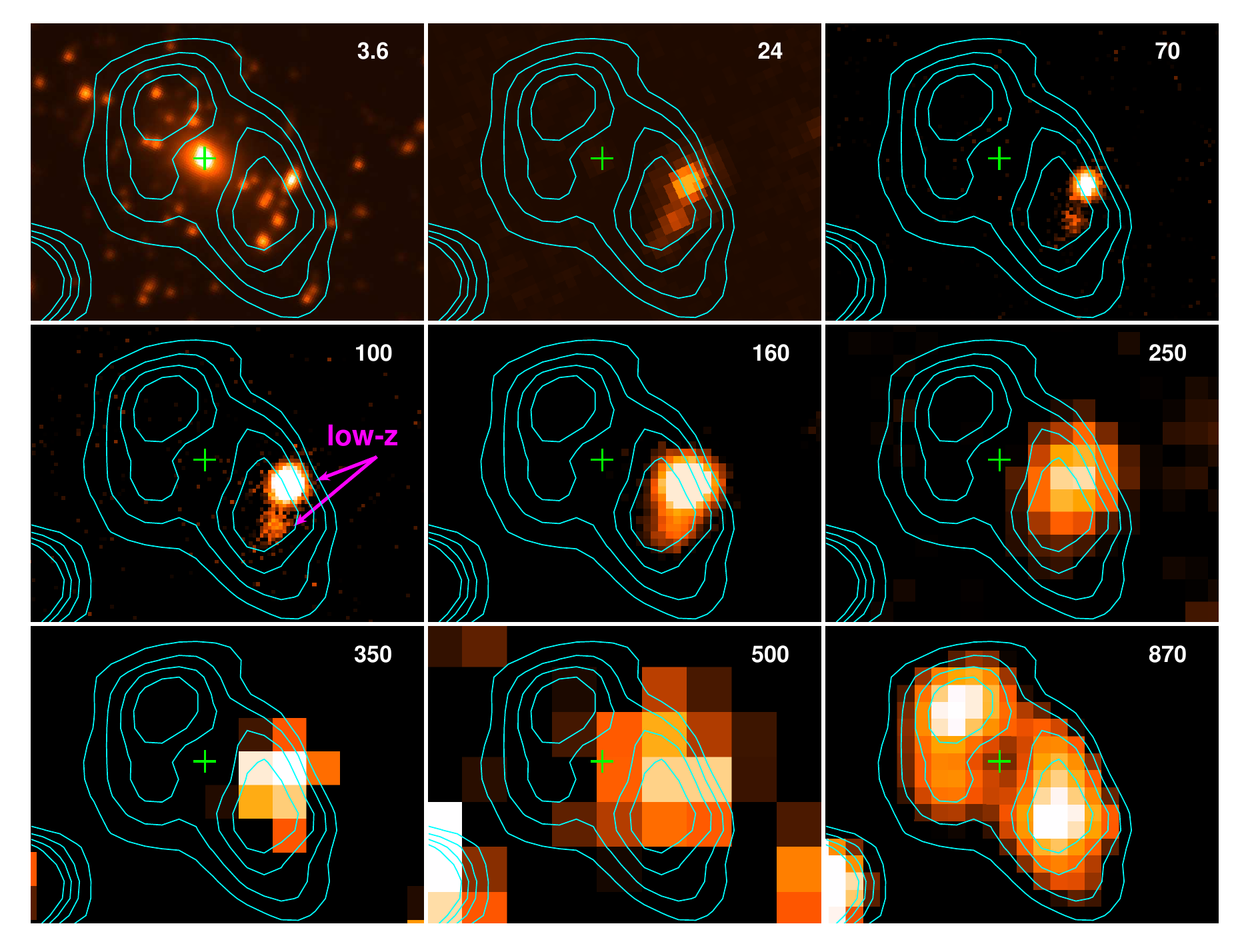}}
\vspace{-5mm}
\caption{
105$''\times$78$''$ thumbnails showing the central region of the cluster AS1063 at 3.6, 24, 70, 100, 160, 250, 350, 500 and  870\,$\mu$m (from left to right and from top to bottom). The contours correspond to the 870\,$\mu$m emission detected with LABOCA at 3, 4, 5 and 6-$\sigma$ ($\sigma=$1.1\,mJy). The {\it Herschel} drop-out source can be seen around the BCG (marked by the green cross) at the center of the 870\,$\mu$m map. 
The arrows in the 100$\mu$m map point two low-z sources ($z=0.$3 and 0.6), whose 870\,$\mu$m emission is blended with the south-western part of the high-z source.}
\label{fig:allw}
\end{figure*}

\begin{figure*}
\centerline{
\includegraphics[width=6.5cm, trim=3mm 11mm 3mm 3mm,clip]{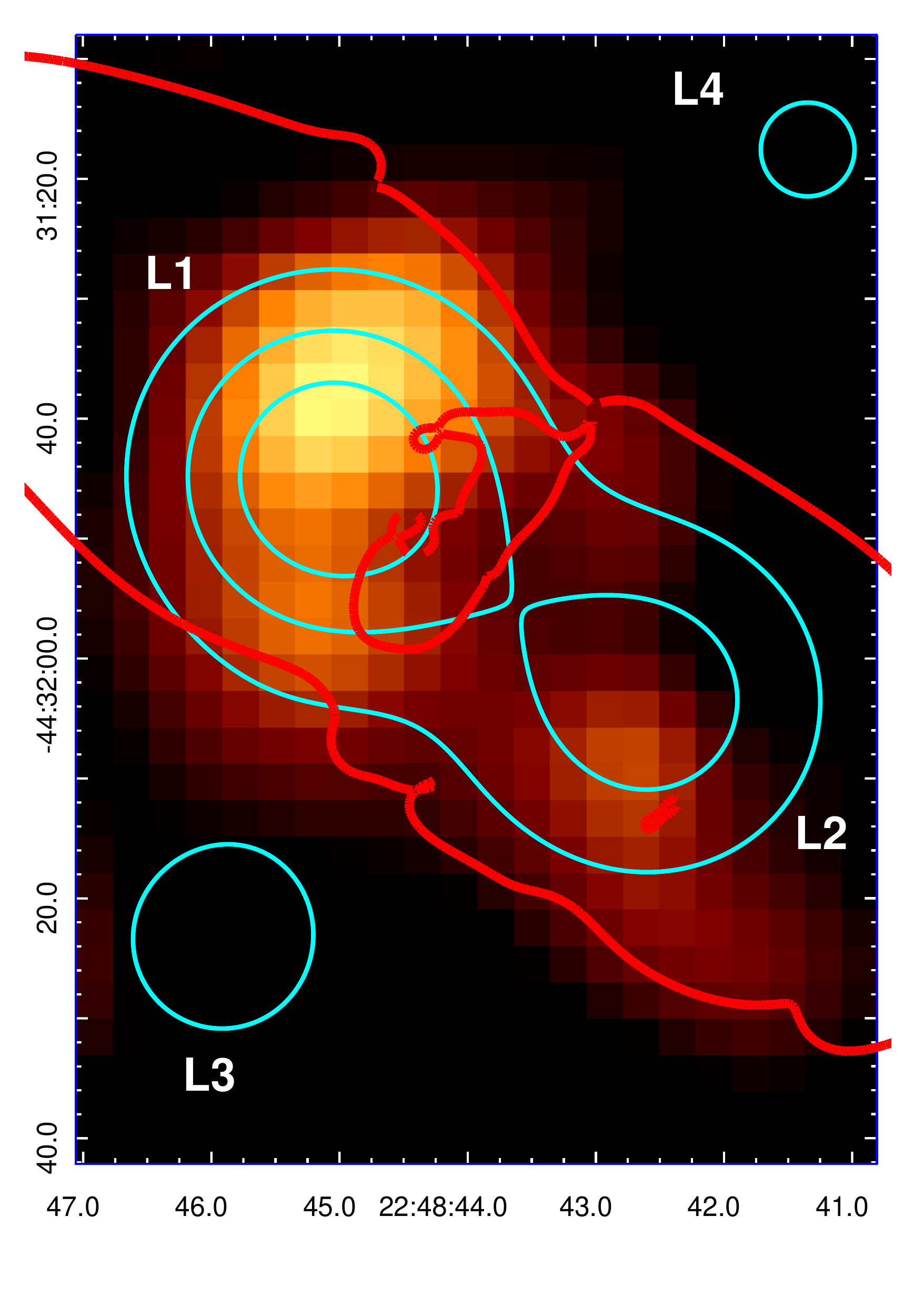}
\includegraphics[width=10.6cm,clip]{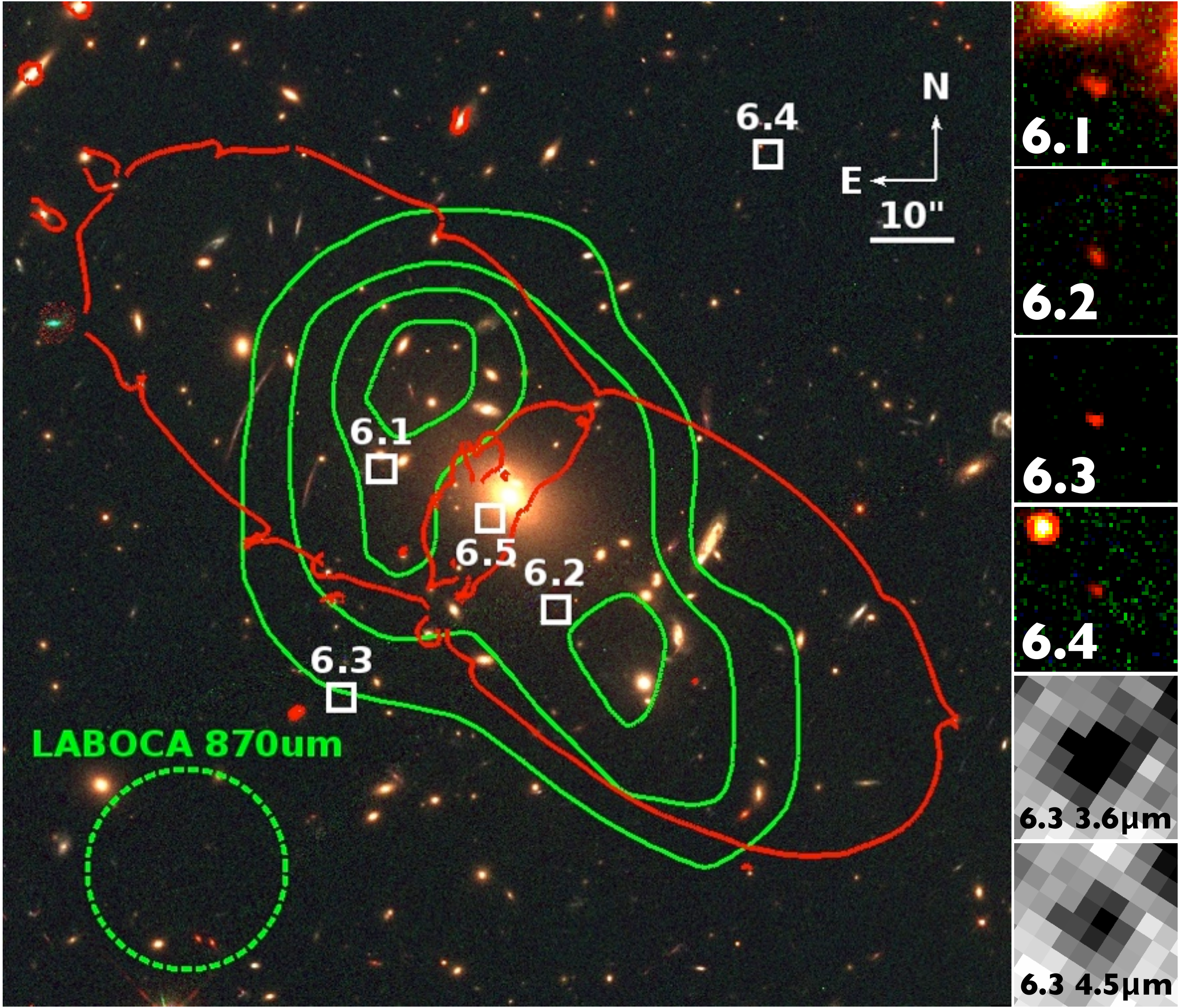}
}
\caption{{\bf Left:} Residuals of the 870$\mu$m emission after subtraction of the two low-z sources. The cyan contours represent the $z=6$ source model lensed by the cluster and observed at the resolution of LABOCA. The four images formed are labeled L1, L2, L3 and L4. The contour levels are at 0.25, 0.5 and 0.75$\times S_{\rm L1}$, where $S_{\rm L1}$ it the peak flux of the L1 image. The critical lines for $z=6$ are overlaid in red. {\bf Right:} {\it HST} color image of the center of the cluster AS1063 made from images in the filters F606W (blue), F775W (green) and F125W (red). The white squares show the positions of the 4 images of a $z=6.1$ background source. The thumbnails on the right are $3''\times 3''$ zooms into these 4 images. The green contours show the 870\,$\mu$m emission at 2.6, 3.9, 5.2 and 6.5\,mJy (RMS=1.1\,mJy). The dotted green circle represents the LABOCA beam (FWHM=24.3$''$). }
\label{fig:clash}
\end{figure*}

\begin{figure*}
\centerline{
\includegraphics[width=18cm,clip]{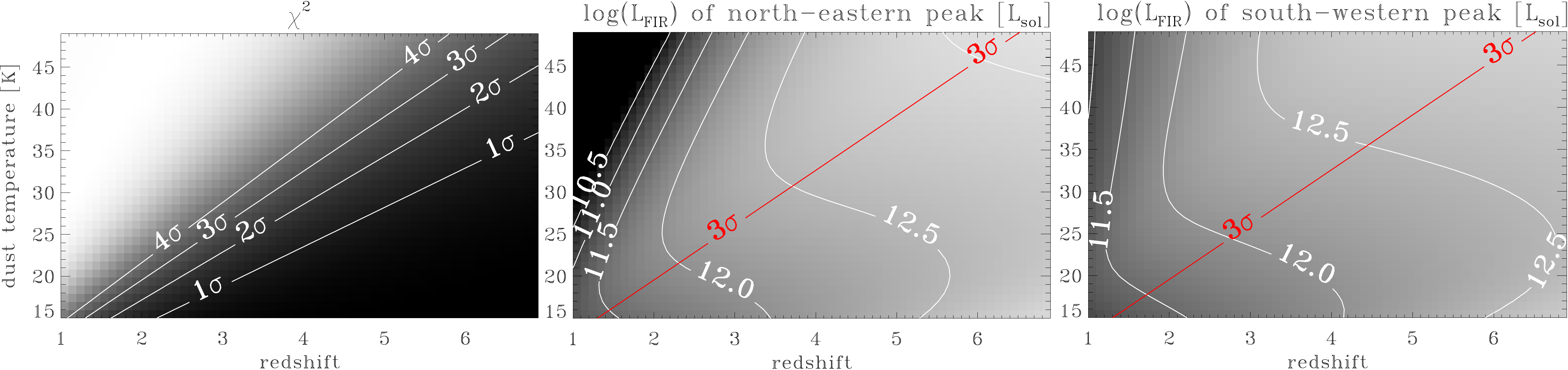}}
\caption{{\bf Left panel:} $\chi^2$ map in the $z$-$T_{\rm d}$ plane, where $z$ and $T_{\rm d}$ are the redshift and the dust temperature assumed for the 870\,$\mu$m source detected with LABOCA. The two  870\,$\mu$m-peaks shown in the Fig.\,\ref{fig:clash} as well as the two low-z sources (at $z=0.6$ and 0.3) shown in the Fig.\,\ref{fig:allw} are fitted simultaneously at all wavelengths from 100 to 870\,$\mu$m assuming a modified black body SED for each source (7 free parameters). The white contours show the 1$\sigma$, 2$\sigma$, 3$\sigma$, 4$\sigma$ confidence levels. 
{\bf Middle and right panels:} the best fit FIR luminosity in log and without any lensing correction in the $z$-$T_{\rm d}$ plane for the two 870\,$\mu$m peaks. The white contours are spaced by 0.5 dex. The $\chi^2$ 3$\sigma$ confidence level is over-plotted in red. 
} 
\label{fig:chi2map}
\end{figure*}

\begin{figure*}
\centerline{
\includegraphics[width=14cm]{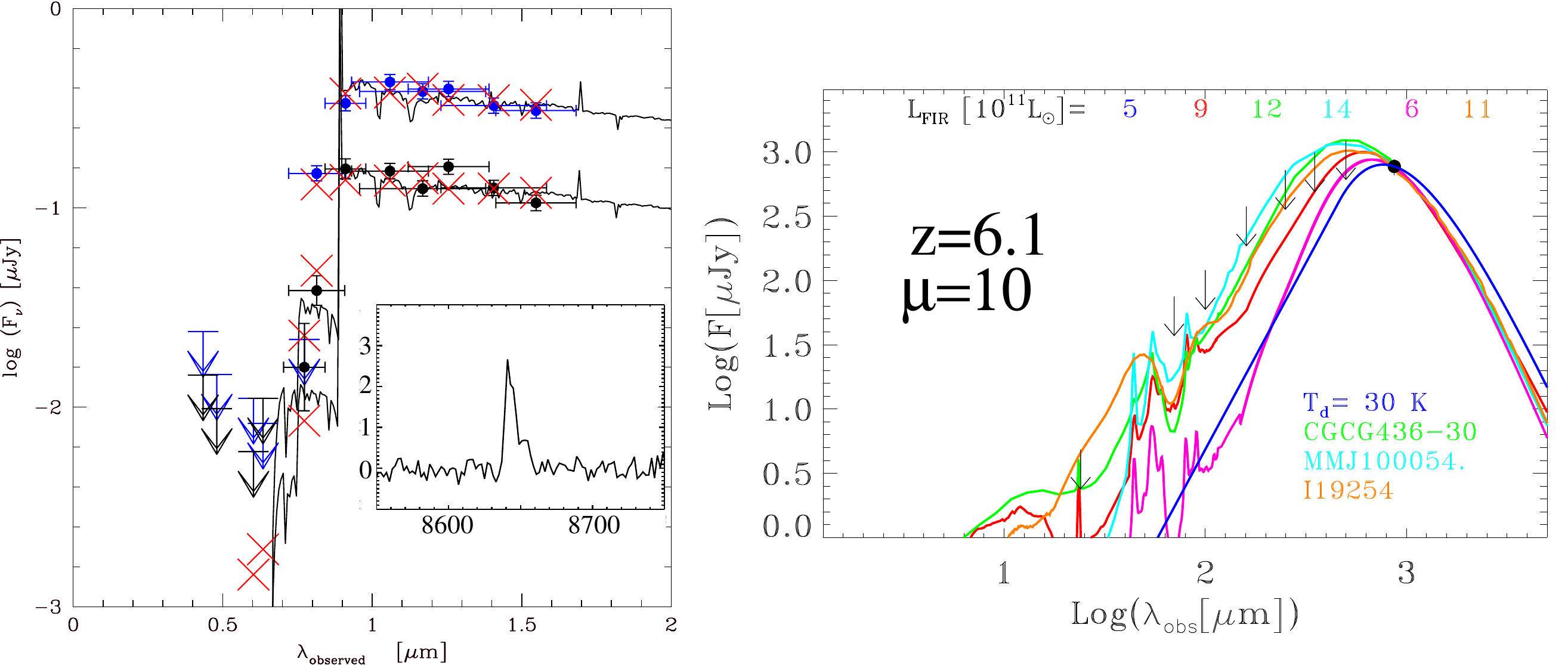}
}
\caption{{\bf Left:} SED fits to the {\it HST}/ACS-WFC3 photometry of the images B and D (A is contaminated by a nearby source, C is in a noisy strip of the detector).  The best fits give $z=6.3\pm 0.3$. The inset shows the VLT/FORS spectrum of the B image with the Ly-$\alpha$ line clearly detected at $z=6.107$ (Richard et al in prep). The x-axis of the inset corresponds to the observed wavelength in $\AA$. {\bf Right:} SED fits to the 870$\mu$m northern peak taking into account the  upper  limits listed in the Table\,\ref{tab:photo} and assuming $z=6.1$ and $\mu=10$ (all the fluxes are corrected for magnification). The modified black body SED with $T_{\rm d}=30$\,K is shown in blue, it gives a FIR luminosity $L_{\rm FIR}=5 \times 10^{11}$\,$L_{\odot}$. The other templates come from the  the \citet{2001ApJ...556..562C} library (red), the \citet{2008A&A...484..631V} library (green), the \citet{2010A&A...514A..67M, 2010ApJ...712..942M} library (cyan) and the \citet{Polletta07} library (orange). The template in magenta corresponds to the SMM\,J2135-0102 model \citep{2010Natur.464..733S,2010A&A...518L..35I}. }
\label{fig:sed}
\end{figure*}

\setlength{\tabcolsep}{3.5pt}
\begin{table}
\caption{Photometry of the 870$\mu$m northern peak with the wavelengths ($\lambda$) in $\mu$m and the flux densities ($S_{\nu}$) in mJy. The upper limts are at 3$\sigma$. }
\vspace{-3mm}
\begin{tabular}{l|cccccccc}
$\lambda$ & 24 & 70 & 100 & 160 & 250 & 350 & 500 & 870 \\
\hline
$S_{\nu}$ & $<$0.06 & $<$0.75 & $<$1.2 & $<$3.8 & $<$7.2 & $<$9.9 & $<$12.3  &  7.6$\pm$1.1\\
\end{tabular} \label{tab:photo}
\vspace{-5mm}
\end{table}

\section{Observations and data reduction}

{\it Herschel} observations of AS1063 at 70, 100, 160, 250, 350 and 500\,$\mu$m  were obtained as part of the {\it Herschel} Lensing Survey (program IDs: KPOT$\_$eegami$\_$1, OT2$\_$trawle$\_$3) as described by \citet{2010A&A...518L..12E} and \citet{2010A&A...518L..14R}. The full width at half maximum (FWHM) of the beams are  5.2$''$, 7.7$''$, 11.3$''$,18.1$''$, 24.9 and 36.6$''$, respectively. 

Observations of AS1063 at $870$\,$\mu$m with the Large APEX Bolometer Camera
LABOCA \citep{2009A&A...497..945S} were obtained in the frame of the LABOCA Lensing Survey, a large program coordinated between ESO and MPI (E187A0437A, M-087.F-0005-2011). The observations were carried out in April and May 2012 in excellent weather conditions with an average precipitable
water vapor (PWV) of 0.5\,mm. The spiral mapping pattern was chosen to cover a circular area of $\sim 8'$ in diameter centered on the clusters. 
Absolute flux calibration was achieved through observations of Mars,
Uranus and Neptune as well as secondary calibrators and 
was found to be accurate within $\sim$10\% (rms). The atmospheric attenuation was
determined via skydips every $\sim$ 2 hours as well as from
independent data from the APEX radiometer which measures the line of
sight water vapor column every minute \citep[see][for a more detailed
description]{2009A&A...497..945S}. Pointing was checked on the nearby
quasars and found to be stable within $3''$ (rms).
The data were reduced using the Bolometer array data Analysis software
(BoA, Schuller \etal\ in prep.).  The { effective resolution of the maps is} $24.3''$. The pixel noise RMS at the center of the map is 1.1\,mJy\,beam$^{-1}$.

\section{Results}
The LABOCA map shows a source at the center of the cluster AS1063 (Fig.\,\ref{fig:allw}) that is extended at the resolution of LABOCA (beam FWHM=24.3$''$) and centered on the brightest cluster galaxy (BCG). Its north-eastern part has no counterparts in any of the {\it Herschel} bands nor  in the 24$\mu$m MIPS band; it is a very red {\it Herschel} drop-out. It peaks at 7.6$\pm$1.1\,mJy and its 3-$\sigma$ upper limit at 500$\mu$m is 13\,mJy, implying a flux ratio $S_{\rm 870}/S_{\rm 500}\ge$0.5.

Although the BCG is not detected with {\it Herschel} \citep{2012ApJ...747...29R}, the south-western part of the 870$\mu$m source is partly blended with the emission coming from two lower redshift sources with spectroscopic redshifts $z=0.6$ and 0.3 (Walth et~al.\ in prep). To extract the source properties in this crowded field we applied a multiwavelength simultaneous fit of the maps assuming a modified black body SED shape \citep[following ][ with $\alpha=2.9$ and $\beta=1.5$]{2003MNRAS.338..733B} for all the sources. There are two free parameters per source corresponding to the FIR luminosity and the wavelength of the SED peak (determined by the dust temperature, $T_{\rm d}$ and the redshift $z$). In a first iteration we run the procedure with the two low-z sources only. The residuals are shown with green contours in the Fig.\,\ref{fig:clash}; they represent the 870$\mu$m foreground-deblended emission. The morphology and the distribution of this emission with respect to the critical lines, with two peaks on each side of the BCG ---one to the northeast at 7.6\,mJy and the other to the southwest at 5.3\,mJy--- {suggest either a multiply imaged background source or substructures in the SZ increment. We discuss the two interpretations in turn in the following section.} 

\section{Discussion}%

\subsection{Photometry of a putative lensed source}
In a second iteration we run the photometry procedure again to fit simultaneously  two sources at the positions of the 870$\mu$m peaks in addition to the two low-z sources. We assume that the two 870$\mu$m sources are the images of a single lensed source, which implies a unique peak wavelength (same redshift and dust temperature) and a total of 7 free parameters. The $\chi^2$ value gives us an indication of the quality of the fit, it is plotted against the redshift and the dust temperature of the lensed source in the Fig.\,\ref{fig:chi2map} with contours indicating the confidence levels. At a given dust temperature, the 870$\mu$m flux and the {\it Herschel} upper limits impose a lower limit to the redshift. Thus, if we assume $T_{\rm d}>20$\,K as observed in most SMGs including the lowest luminosities \citep[see, e.g.,][or Swinbank et al.\ submitted]{2013MNRAS.431.2317S,2012A&A...539A.155M}, then  the source must be at $z\ge 2$. If  $T_{\rm d}=30$\,K  { (mean value for  $L_{\rm FIR}\sim10^{11.5}L_{\odot}$ according to the same references)}, then the redshift must be $\ge 4$. In addition, if we assume $z<7$, the observed (i.e., uncorrected for lensing)  FIR luminosities of the two peaks must be $<10^{13}L_{\odot}$ (Fig.\,\ref{fig:chi2map}). 

\subsection{Lens and source models}

Based on the identification of 13 multiple image systems out of which 5 have a spectroscopic redshift and the others have a reliable photometric redshift, we have built a lens model of the cluster (Richard et al.\ in prep). The critical lines computed with this model for $z$=6 are shown in the  Fig.\,\ref{fig:clash} in red. 

As shown on the left panel of Fig.\,\ref{fig:clash}, with this model we can reproduce the two 870$\mu$m peaks by assuming a single source modeled by a circular Gaussian of FWHM=2\,kpc  at $z=6$.
Four images of the source, labeled L1, L2, L3 and L4, are actually formed in a classical quad configuration; their magnifications are 10.8, 3.7, 7.1 and 3.1, respectively for a total magnification $\mu$=25.5. The images L3 and L4 are $\sim$3$\times$ fainter than L1 and are therefore at $\sim 2\sigma$, which is consistent with no detection. To obtain an L1-image brighter than the others it is required to align it with a galaxy of the cluster such that it undergoes an additional magnification. In this model the L1-image is formed close to two galaxies of the cluster. 
The differences between the model and the data are $\le$3$\sigma$. Because the angular distance between the two peaks (L1 and L2) decreases with decreasing redshift our lens model puts a strong constraint on the redshift, which must be $\ge 4$ if we assume a single source. { We note that it is possible that the southern source arises from lensing of a second background galaxy. However, this would imply multiple sources with similar very red SEDs.}

Hence, according to our lens$+$source model and to the photometry, the luminosity of the { putative} lensed source corrected for lensing is likely $< 10^{12} L_{\odot}$, i.e., an order of magnitude lower than that of SMGs detected at $z>4$ so far.
For example, if we assume $T_{\rm d}=30$\,K  and $z=6$, the observed luminosity of the northern peak is $L_{\rm FIR}\sim 5 \times 10^{12}L_{\odot}$ (middle panel of  Fig.\,\ref{fig:chi2map}), with $\mu_{\rm L1}=10$ this implies an intrinsic luminosity $L_{\rm FIR}\sim 5\times 10^{11}L_{\odot}$. 

\subsection{A plausible {\it HST} counterpart at $z=6.107$}
In the {\it HST} images and catalogs provided by the CLASH project we identified 4 objects (named 6.1, 6.2, 6.3 and 6.4 in Fig.\,\ref{fig:clash} as in Richard et al. in prep), which could be the 4 images of a high-z source. Indeed, fitting various SED templates to the HST photometry we derived a redshift  $z=6.3\pm 0.3$ (Fig\,\ref{fig:sed}) and the image positions were accurately reproduced by our lens model for a source at $z\sim 6$.  To confirm the redshift we recently obtained VLT/FORS spectroscopy of the 6.2, 6.3 and 6.4 images. The Ly-$\alpha$ line  is clearly detected at  $z=6.107$ (Fig.\,\ref{fig:sed})
\footnote{
In a recent paper \citet{2013arXiv1308.1692B} mention a quintuple system in this cluster but they list only 3 of our identified images (6.1, 6.3 and 6.4). After the submission of our letter two other articles appeared online \citep{2013arXiv1308.6280M, 2013arXiv1309.1593B} describing in more detail this system and providing a similar spectroscopic redshift. We confirm the fifth image identified by these authors and show it in the Fig.\,\ref{fig:clash} as "6.5".}. 
The magnifications are $\mu_{6.1}$=17.1, $\mu_{6.2}$=6.7, $\mu_{6.3}$=5.9 and $\mu_{6.4}$=2.5.

The image 6.1 of this {\it HST} source benefits from a boost by the same two galaxies of the cluster as in the model discussed above for the 870$\mu$m emission. However, if they are both at the same redshift the LABOCA source needs to be offset from the {\it HST} source to reproduce the southern peak (L2); it is at $\sim$30\,kpc in the  above model. The distance between the two sources is mainly constrained by the flux ratio of the two 870$\mu$m peaks, it could be in the range 10-30\,kpc, suggesting interacting galaxies. 

The star formation rate (SFR) of the {\it HST} source estimated from the UV continuum and from the Ly-$\alpha$ line and corrected for lensing are SFR$_{\rm UV}\sim 5$\,$M_{\odot}$\,yr$^{-1}$ and SFR$_{Ly\alpha}\sim 15$\,$M_{\odot}$\,yr$^{-1}$. The {\it Spitzer} detection of 6.3 at 3.6$\mu$m implies  (H-3.6) $\approx 2$,   redder than for typical $z \sim$ 6--8 LBGs \citep{2011MNRAS.418.2074M}. SED fits with young populations predict up to $A_v\sim 1.5$ implying a reprocessed IR luminosity of $\sim 4\times 10^{11}L_{\odot}$, i.e., an SFR$_{\rm IR}\sim 70$\,$M_{\odot}$\,yr$^{-1}$.

At $z=6.1$ the upper limit on the FIR luminosity of the 870$\mu$m source depends on the SED template  assumed as illustrated on the right panel of Fig.\,\ref{fig:sed}. 
The intrinsic luminosities obtained are in the range $[5-15]\times10^{11}L_{\odot}$, which corresponds to an SFR in the range $[80-260]$\,$M_{\odot}$\,yr$^{-1}$. The star forming properties of the  {\it HST} and the LABOCA sources could therefore be similar.
\subsection{SZ substructure in the merging cluster?}

\begin{figure}
\centerline{
\includegraphics[width=7cm]{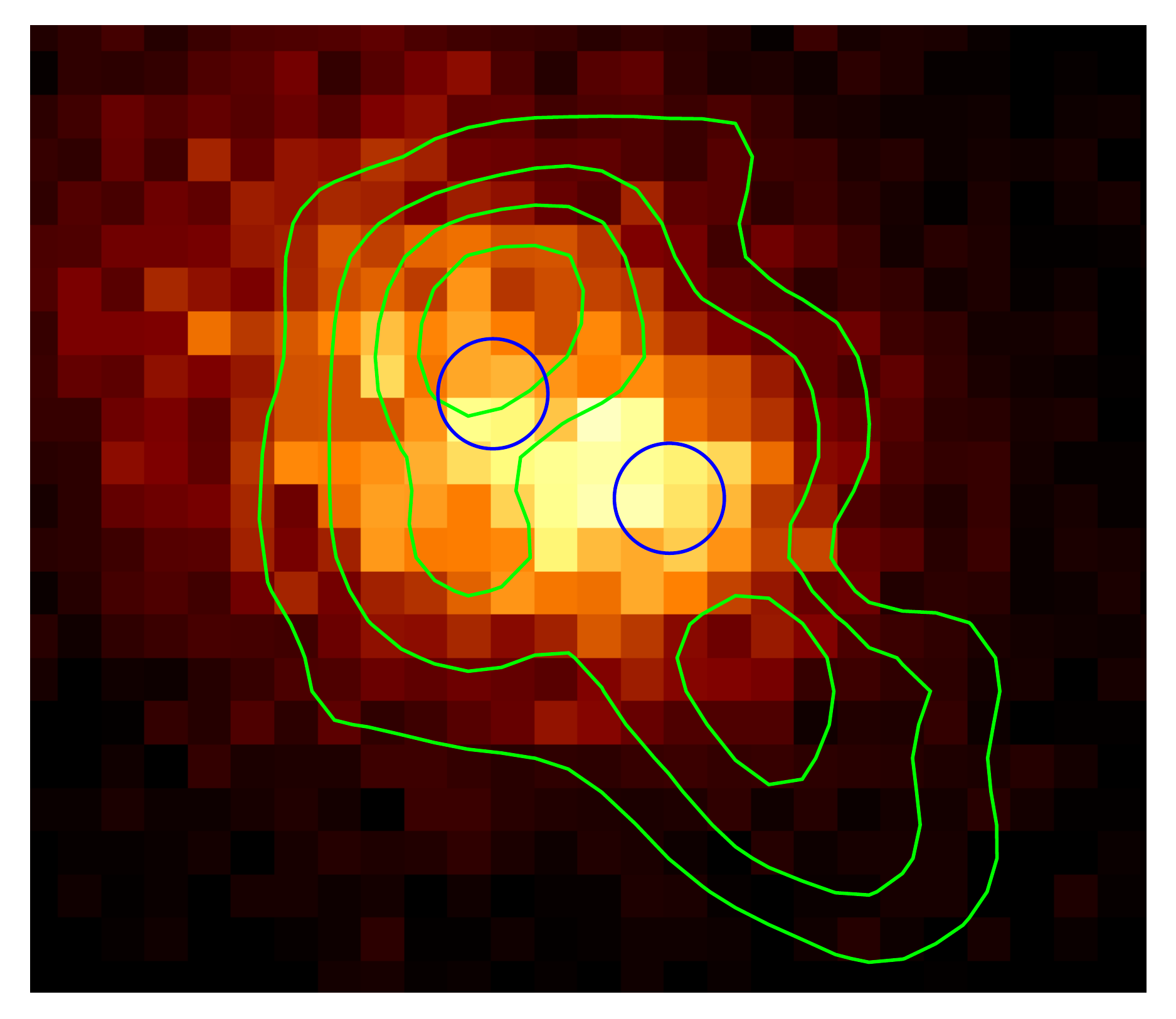}
}
\vspace*{-3mm}
\caption{ Chandra X-ray map of AS1063 overlaid with the LABOCA residuals in green contours starting at 2$\sigma$ and spaced by 1$\sigma$. The blue circles mark the positions the two components of the $\beta$-model fitted by \citet{2012AJ....144...79G} to the Chandra data. }
\label{fig:chandra}
\end{figure}

AS1063 is known to yield a strong SZ effect \citep{2010ApJ...716.1118P} that was detected at S/N$\sim$17 with {\it Planck} \citep{2013arXiv1303.5089P}.  Furthermore, \citet{2012AJ....144...79G} showed that this cluster is undergoing a major merging event close to the plane of the sky and there has been growing evidences in recent years, both from observations and simulations, that such a merging configuration can produce small scale substructures in the SZ \citep{2011ApJ...734...10K, 2012ApJ...761...47M,2013MNRAS.432.3508R}. These substructures could be caused by shocks and inhomogeneities in the hot gas and their increment could peak at 300-400\,GHz \citep[i.e., 750-1000\,$\mu$m, ][]{2013MNRAS.432.3508R}. The SED of such SZ substructures could therefore be consistent with the {\it Herschel} and LABOCA photometry. Also we note that the northern LABOCA peak is close to the secondary mass component identified by \citet{2012AJ....144...79G} (Fig.\,\ref{fig:chandra}) and its elongated morphology would be consistent with shocked gas. The large scale SZ would be filtered out by the LABOCA observations and the data reduction, and we would be seeing the substructures related to the merging event. { More quantitatively, the {\it Planck} measurement at 353\,GHz (the central pixel is at 0.12\,MJy/sr) can be used to scale the SZ profile modeled by \citet{2010ApJ...716.1118P} for this cluster and  thus estimate the peak flux expected at the center of the LABOCA map ignoring spatial filtering. We obtain $\sim$7\,mJy/beam. The fraction of this flux filtered by the observations is difficult to estimate from the data because it depends on the morphology of the SZ increment.} This will be studied in a forthcoming paper by Zemcov et al.\ (in prep). 

\section{Conclusion}

With APEX/LABOCA we have detected an extended $870$\,$\mu$m source aligned with the center of the cluster AS1063. The source is not detected at shorter FIR/submm wavelengths. We find two possible interpretations of this peculiar source, it could be the dusty component of an HST-detected strongly-lensed galaxy at $z$=6.1 or substructures in the SZ effect. The current observations do not allow us to disentangle the two interpretations.

There are two routes to decide between the different origins of the features we have discovered: submm observations with ALMA should allow us to resolve the four images of the high-z source, while lower frequency observations (150\,GHz) are required to measure the decrement of the SZ substructures.

\begin{acknowledgements}
We are very grateful to the APEX staff for their great help with the observations and their nice welcome at the Sequitor base. We gratefully aknowledge the ESO director for the VLT/FORS program DDT 291.A-5027. We kindly acknowledge Etienne Pointecouteau for providing us with the {\it Planck} data and for useful discussions. This work received support from the Agence Nationale de la Recherche bearing the reference ANR-09-BLAN-0234. JPK thanks support from the European Research Council (ERC) advanced grant “Light on the Dark” (LIDA) and CNRS. IRS acknowledges support from STFC (ST/I001573/1), a Leverhulme Fellowship, the ERC Advanced Investigator programme DUSTYGAL 321334 and a Royal Society/Wolfson Merit Award. AMS acknowledges an STFC Advanced Fellowship through grant ST/H005234/1. KK thanks the Swedish Research Council for support (grant 621-2011-5372).
\end{acknowledgements}

\bibliographystyle{aa}
\bibliography{hls-laboca,hls_highz,sz}

\begin{thebibliography}{36}
\expandafter\ifx\csname natexlab\endcsname\relax\def\natexlab#1{#1}\fi

\bibitem[{{Balestra} {et~al.}(2013){Balestra}, {Vanzella}, {Rosati}, {Monna},
  {Grillo}, {Nonino}, {Mercurio}, {Biviano}, {Bradley}, {Coe}, {Fritz},
  {Postman}, {Seitz}, {Scodeggio}, {Zheng}, {Ziegler}, {Zitrin},
  {Annunziatella}, {Bartelmann}, {Benitez}, {Broadhurst}, {Bouwens}, {Czoske},
  {Donahue}, {Ford}, {Girardi}, {Infante}, {Kelson}, {Koekemoer}, {Kuchner},
  {Lemze}, {Lombardi}, {Maier}, {Medezinski}, {Melchior}, {Meneghetti},
  {Merten}, {Moustakas}, {Presotto}, {Smit}, {Tozzi}, \&
  {Umetsu}}]{2013arXiv1309.1593B}
{Balestra}, I., {Vanzella}, E., {Rosati}, P., {et~al.} 2013, arXiv/1309.1593

\bibitem[{{Blain} {et~al.}(2003){Blain}, {Barnard}, \&
  {Chapman}}]{2003MNRAS.338..733B}
{Blain}, A.~W., {Barnard}, V.~E., \& {Chapman}, S.~C. 2003, \mnras, 338, 733

\bibitem[{{Blain} {et~al.}(2002){Blain}, {Smail}, {Ivison}, {Kneib}, \&
  {Frayer}}]{2002PhR...369..111B}
{Blain}, A.~W., {Smail}, I., {Ivison}, R.~J., {Kneib}, J.-P., \& {Frayer},
  D.~T. 2002, \physrep, 369, 111

\bibitem[{{Bradley} {et~al.}(2013){Bradley}, {Zitrin}, {Coe}, {Bouwens},
  {Postman}, {Balestra}, {Grillo}, {Monna}, {Rosati}, {Seitz}, {Host}, {Lemze},
  {Moustakas}, {Moustakas}, {Shu}, {Zheng}, {Broadhurst}, {Carrasco}, {Jouvel},
  {Koekemoer}, {.~Medezinski}, {Meneghetti}, {Nonino}, {Smit}, {Umetsu},
  {Bartelmann}, {Benitez}, {Donahue}, {Ford}, {Infante}, {Jimenez-Teja},
  {Kelson}, {Lahav}, {Maoz}, {Melchior}, {Merten}, \&
  {Molino}}]{2013arXiv1308.1692B}
{Bradley}, L.~D., {Zitrin}, A., {Coe}, D., {et~al.} 2013, arXiv/1308.1692

\bibitem[{{Capak} {et~al.}(2011){Capak}, {Riechers}, {Scoville}, {Carilli},
  {Cox}, {Neri}, {Robertson}, {Salvato}, {Schinnerer}, {Yan}, {Wilson}, {Yun},
  {Civano}, {Elvis}, {Karim}, {Mobasher}, \& {Staguhn}}]{2011Natur.470..233C}
{Capak}, P.~L., {Riechers}, D., {Scoville}, N.~Z., {et~al.} 2011, \nat, 470,
  233

\bibitem[{{Chapman} {et~al.}(2005){Chapman}, {Blain}, {Smail}, \&
  {Ivison}}]{2005ApJ...622..772C}
{Chapman}, S.~C., {Blain}, A.~W., {Smail}, I., \& {Ivison}, R.~J. 2005, \apj,
  622, 772

\bibitem[{{Chary} \& {Elbaz}(2001)}]{2001ApJ...556..562C}
{Chary}, R. \& {Elbaz}, D. 2001, \apj, 556, 562

\bibitem[{{Combes} {et~al.}(2012){Combes}, {Rex}, {Rawle}, {Egami}, {Boone},
  {Smail}, {Richard}, {Ivison}, {Gurwell}, {Casey}, {Omont}, {Berciano Alba},
  {Dessauges-Zavadsky}, {Edge}, {Fazio}, {Kneib}, {Okabe}, {Pell{\'o}},
  {P{\'e}rez-Gonz{\'a}lez}, {Schaerer}, {Smith}, {Swinbank}, \& {van der
  Werf}}]{2012A&A...538L...4C}
{Combes}, F., {Rex}, M., {Rawle}, T.~D., {et~al.} 2012, \aap, 538, L4

\bibitem[{{Egami} {et~al.}(2010){Egami}, {Rex}, {Rawle},
  {P{\'e}rez-Gonz{\'a}lez}, {Richard}, {Kneib}, {Schaerer}, {Altieri},
  {Valtchanov}, {Blain}, {Fadda}, {Zemcov}, {Bock}, {Boone}, {Bridge},
  {Clement}, {Combes}, {Dessauges-Zavadsky}, {Dowell}, {Ilbert}, {Ivison},
  {Jauzac}, {Lutz}, {Metcalfe}, {Omont}, {Pell{\'o}}, {Pereira}, {Rieke},
  {Rodighiero}, {Smail}, {Smith}, {Tramoy}, {Walth}, {van der Werf}, \&
  {Werner}}]{2010A&A...518L..12E}
{Egami}, E., {Rex}, M., {Rawle}, T.~D., {et~al.} 2010, \aap, 518, L12+

\bibitem[{{G{\'o}mez} {et~al.}(2012){G{\'o}mez}, {Valkonen}, {Romer},
  {Lloyd-Davies}, {Verdugo}, {Cantalupo}, {Daub}, {Goldstein}, {Kuo}, {Lange},
  {Lueker}, {Holzapfel}, {Peterson}, {Ruhl}, {Runyan}, {Reichardt}, \&
  {Sabirli}}]{2012AJ....144...79G}
{G{\'o}mez}, P.~L., {Valkonen}, L.~E., {Romer}, A.~K., {et~al.} 2012, \aj, 144,
  79

\bibitem[{{Hatsukade} {et~al.}(2013){Hatsukade}, {Ohta}, {Seko}, {Yabe}, \&
  {Akiyama}}]{2013ApJ...769L..27H}
{Hatsukade}, B., {Ohta}, K., {Seko}, A., {Yabe}, K., \& {Akiyama}, M. 2013,
  \apjl, 769, L27

\bibitem[{{Ivison} {et~al.}(2010){Ivison}, {Swinbank}, {Swinyard}, {Smail},
  {Pearson}, {Rigopoulou}, {Polehampton}, {Baluteau}, {Barlow}, {Blain},
  {Bock}, {Clements}, {Coppin}, {Cooray}, {Danielson}, {Dwek}, {Edge},
  {Franceschini}, {Fulton}, {Glenn}, {Griffin}, {Isaak}, {Leeks}, {Lim},
  {Naylor}, {Oliver}, {Page}, {P{\'e}rez Fournon}, {Rowan-Robinson}, {Savini},
  {Scott}, {Spencer}, {Valtchanov}, {Vigroux}, \&
  {Wright}}]{2010A&A...518L..35I}
{Ivison}, R.~J., {Swinbank}, A.~M., {Swinyard}, B., {et~al.} 2010, \aap, 518,
  L35+

\bibitem[{{Karim} {et~al.}(2013){Karim}, {Swinbank}, {Hodge}, {Smail},
  {Walter}, {Biggs}, {Simpson}, {Danielson}, {Alexander}, {Bertoldi}, {de
  Breuck}, {Chapman}, {Coppin}, {Dannerbauer}, {Edge}, {Greve}, {Ivison},
  {Knudsen}, {Menten}, {Schinnerer}, {Wardlow}, {Wei{\ss}}, \& {van der
  Werf}}]{2013MNRAS.432....2K}
{Karim}, A., {Swinbank}, A.~M., {Hodge}, J.~A., {et~al.} 2013, \mnras, 432, 2

\bibitem[{{Knudsen} {et~al.}(2010){Knudsen}, {Kneib}, {Richard}, {Petitpas}, \&
  {Egami}}]{2010ApJ...709..210K}
{Knudsen}, K.~K., {Kneib}, J.-P., {Richard}, J., {Petitpas}, G., \& {Egami}, E.
  2010, \apj, 709, 210

\bibitem[{{Korngut} {et~al.}(2011){Korngut}, {Dicker}, {Reese}, {Mason},
  {Devlin}, {Mroczkowski}, {Sarazin}, {Sun}, \&
  {Sievers}}]{2011ApJ...734...10K}
{Korngut}, P.~M., {Dicker}, S.~R., {Reese}, E.~D., {et~al.} 2011, \apj, 734, 10

\bibitem[{{Magnelli} {et~al.}(2012){Magnelli}, {Lutz}, {Santini}, {Saintonge},
  {Berta}, {Albrecht}, {Altieri}, {Andreani}, {Aussel}, {Bertoldi},
  {B{\'e}thermin}, {Bongiovanni}, {Capak}, {Chapman}, {Cepa}, {Cimatti},
  {Cooray}, {Daddi}, {Danielson}, {Dannerbauer}, {Dunlop}, {Elbaz}, {Farrah},
  {F{\"o}rster Schreiber}, {Genzel}, {Hwang}, {Ibar}, {Ivison}, {Le Floc'h},
  {Magdis}, {Maiolino}, {Nordon}, {Oliver}, {P{\'e}rez Garc{\'{\i}}a},
  {Poglitsch}, {Popesso}, {Pozzi}, {Riguccini}, {Rodighiero}, {Rosario},
  {Roseboom}, {Salvato}, {Sanchez-Portal}, {Scott}, {Smail}, {Sturm},
  {Swinbank}, {Tacconi}, {Valtchanov}, {Wang}, \&
  {Wuyts}}]{2012A&A...539A.155M}
{Magnelli}, B., {Lutz}, D., {Santini}, P., {et~al.} 2012, \aap, 539, A155

\bibitem[{{McLure} {et~al.}(2011){McLure}, {Dunlop}, {de Ravel}, {Cirasuolo},
  {Ellis}, {Schenker}, {Robertson}, {Koekemoer}, {Stark}, \&
  {Bowler}}]{2011MNRAS.418.2074M}
{McLure}, R.~J., {Dunlop}, J.~S., {de Ravel}, L., {et~al.} 2011, \mnras, 418,
  2074

\bibitem[{{Micha{\l}owski} {et~al.}(2010{\natexlab{a}}){Micha{\l}owski},
  {Hjorth}, \& {Watson}}]{2010A&A...514A..67M}
{Micha{\l}owski}, M., {Hjorth}, J., \& {Watson}, D. 2010{\natexlab{a}}, \aap,
  514, A67+

\bibitem[{{Micha{\l}owski} {et~al.}(2010{\natexlab{b}}){Micha{\l}owski},
  {Watson}, \& {Hjorth}}]{2010ApJ...712..942M}
{Micha{\l}owski}, M.~J., {Watson}, D., \& {Hjorth}, J. 2010{\natexlab{b}},
  \apj, 712, 942

\bibitem[{{Monna} {et~al.}(2013){Monna}, {Seitz}, {Greisel}, {Eichner},
  {Drory}, {Postman}, {Zitrin}, {Coe}, {Halkola}, {Suyu}, {Grillo}, {Rosati},
  {Lemze}, {Balestra}, {Snigula}, {Bradley}, {Umetsu}, {Koekemoer},
  {Bartelmann}, {Benitez}, {Bouwens}, {Broadhurst}, {Donahue}, {Ford}, {Host},
  {Infante}, {Jimenez-Teja}, {Jouvel}, {Kelson}, {Lahav}, {Medezinski},
  {Melchior}, {Meneghetti}, {Merten}, {Molino}, {Moustakas}, {Moustakas},
  {Nonino}, \& {Zheng}}]{2013arXiv1308.6280M}
{Monna}, A., {Seitz}, S., {Greisel}, N., {et~al.} 2013, arXiv/1308.6280

\bibitem[{{Mroczkowski} {et~al.}(2012){Mroczkowski}, {Dicker}, {Sayers},
  {Reese}, {Mason}, {Czakon}, {Romero}, {Young}, {Devlin}, {Golwala},
  {Korngut}, {Sarazin}, {Bock}, {Koch}, {Lin}, {Molnar}, {Pierpaoli}, {Umetsu},
  \& {Zemcov}}]{2012ApJ...761...47M}
{Mroczkowski}, T., {Dicker}, S., {Sayers}, J., {et~al.} 2012, \apj, 761, 47

\bibitem[{{Plagge} {et~al.}(2010){Plagge}, {Benson}, {Ade}, {Aird}, {Bleem},
  {Carlstrom}, {Chang}, {Cho}, {Crawford}, {Crites}, {de Haan}, {Dobbs},
  {George}, {Hall}, {Halverson}, {Holder}, {Holzapfel}, {Hrubes}, {Joy},
  {Keisler}, {Knox}, {Lee}, {Leitch}, {Lueker}, {Marrone}, {McMahon}, {Mehl},
  {Meyer}, {Mohr}, {Montroy}, {Padin}, {Pryke}, {Reichardt}, {Ruhl},
  {Schaffer}, {Shaw}, {Shirokoff}, {Spieler}, {Stalder}, {Staniszewski},
  {Stark}, {Vanderlinde}, {Vieira}, {Williamson}, \&
  {Zahn}}]{2010ApJ...716.1118P}
{Plagge}, T., {Benson}, B.~A., {Ade}, P.~A.~R., {et~al.} 2010, \apj, 716, 1118

\bibitem[{{Planck Collaboration} {et~al.}(2013){Planck Collaboration}, {Ade},
  {Aghanim}, {Armitage-Caplan}, {Arnaud}, {Ashdown}, {Atrio-Barandela},
  {Aumont}, {Aussel}, {Baccigalupi}, \& et~al.}]{2013arXiv1303.5089P}
{Planck Collaboration}, {Ade}, P.~A.~R., {Aghanim}, N., {et~al.} 2013,
  arXiv/1303.5089

\bibitem[{Polletta {et~al.}(2007)Polletta, Tajer, Maraschi, Trinchieri,
  Lonsdale, Chiappetti, Andreon, Pierre, Le~F{\`e}vre, Zamorani, Maccagni,
  Garcet, Surdej, Franceschini, Alloin, Shupe, Surace, Fang, Rowan-Robinson,
  Smith, \& Tresse}]{Polletta07}
Polletta, M., Tajer, M., Maraschi, L., {et~al.} 2007, \apj, 81

\bibitem[{{Rawle} {et~al.}(2010){Rawle}, {Chung}, {Fadda}, {Rex}, {Egami},
  {P{\'e}rez-Gonz{\'a}lez}, {Altieri}, {Blain}, {Bridge}, {Fiedler},
  {Gonzalez}, {Pereira}, {Richard}, {Smail}, {Valtchanov}, {Zemcov},
  {Appleton}, {Bock}, {Boone}, {Clement}, {Combes}, {Dowell},
  {Dessauges-Zavadsky}, {Ilbert}, {Ivison}, {Jauzac}, {Kneib}, {Lutz},
  {Pell{\'o}}, {Rieke}, {Rodighiero}, {Schaerer}, {Smith}, {Walth}, {van der
  Werf}, \& {Werner}}]{2010A&A...518L..14R}
{Rawle}, T.~D., {Chung}, S.~M., {Fadda}, D., {et~al.} 2010, \aap, 518, L14+

\bibitem[{{Rawle} {et~al.}(2012){Rawle}, {Edge}, {Egami}, {Rex}, {Smith},
  {Altieri}, {Fiedler}, {Haines}, {Pereira}, {P{\'e}rez-Gonz{\'a}lez},
  {Portouw}, {Valtchanov}, {Walth}, {van der Werf}, \&
  {Zemcov}}]{2012ApJ...747...29R}
{Rawle}, T.~D., {Edge}, A.~C., {Egami}, E., {et~al.} 2012, \apj, 747, 29

\bibitem[{{Riechers} {et~al.}(2013){Riechers}, {Bradford}, {Clements},
  {Dowell}, {P{\'e}rez-Fournon}, {Ivison}, {Bridge}, {Conley}, {Fu}, {Vieira},
  {Wardlow}, {Calanog}, {Cooray}, {Hurley}, {Neri}, {Kamenetzky}, {Aguirre},
  {Altieri}, {Arumugam}, {Benford}, {B{\'e}thermin}, {Bock}, {Burgarella},
  {Cabrera-Lavers}, {Chapman}, {Cox}, {Dunlop}, {Earle}, {Farrah}, {Ferrero},
  {Franceschini}, {Gavazzi}, {Glenn}, {Solares}, {Gurwell}, {Halpern},
  {Hatziminaoglou}, {Hyde}, {Ibar}, {Kov{\'a}cs}, {Krips}, {Lupu}, {Maloney},
  {Martinez-Navajas}, {Matsuhara}, {Murphy}, {Naylor}, {Nguyen}, {Oliver},
  {Omont}, {Page}, {Petitpas}, {Rangwala}, {Roseboom}, {Scott}, {Smith},
  {Staguhn}, {Streblyanska}, {Thomson}, {Valtchanov}, {Viero}, {Wang},
  {Zemcov}, \& {Zmuidzinas}}]{2013Natur.496..329R}
{Riechers}, D.~A., {Bradford}, C.~M., {Clements}, D.~L., {et~al.} 2013, \nat,
  496, 329

\bibitem[{{Ruan} {et~al.}(2013){Ruan}, {Quinn}, \&
  {Babul}}]{2013MNRAS.432.3508R}
{Ruan}, J.~J., {Quinn}, T.~R., \& {Babul}, A. 2013, \mnras, 432, 3508

\bibitem[{{Siringo} {et~al.}(2009){Siringo}, {Kreysa}, {Kov{\'a}cs},
  {Schuller}, {Wei{\ss}}, {Esch}, {Gem{\"u}nd}, {Jethava}, {Lundershausen},
  {Colin}, {G{\"u}sten}, {Menten}, {Beelen}, {Bertoldi}, {Beeman}, \&
  {Haller}}]{2009A&A...497..945S}
{Siringo}, G., {Kreysa}, E., {Kov{\'a}cs}, A., {et~al.} 2009, \aap, 497, 945

\bibitem[{{Smail} {et~al.}(1997){Smail}, {Ivison}, \&
  {Blain}}]{1997ApJ...490L...5S}
{Smail}, I., {Ivison}, R.~J., \& {Blain}, A.~W. 1997, \apjl, 490, L5

\bibitem[{{Swinbank} {et~al.}(2010){Swinbank}, {Smail}, {Longmore}, {Harris},
  {Baker}, {De Breuck}, {Richard}, {Edge}, {Ivison}, {Blundell}, {Coppin},
  {Cox}, {Gurwell}, {Hainline}, {Krips}, {Lundgren}, {Neri}, {Siana},
  {Siringo}, {Stark}, {Wilner}, \& {Younger}}]{2010Natur.464..733S}
{Swinbank}, A.~M., {Smail}, I., {Longmore}, S., {et~al.} 2010, \nat, 464, 733

\bibitem[{{Symeonidis} {et~al.}(2013){Symeonidis}, {Vaccari}, {Berta}, {Page},
  {Lutz}, {Arumugam}, {Aussel}, {Bock}, {Boselli}, {Buat}, {Capak}, {Clements},
  {Conley}, {Conversi}, {Cooray}, {Dowell}, {Farrah}, {Franceschini},
  {Giovannoli}, {Glenn}, {Griffin}, {Hatziminaoglou}, {Hwang}, {Ibar},
  {Ilbert}, {Ivison}, {Floc'h}, {Lilly}, {Kartaltepe}, {Magnelli}, {Magdis},
  {Marchetti}, {Nguyen}, {Nordon}, {O'Halloran}, {Oliver}, {Omont},
  {Papageorgiou}, {Patel}, {Pearson}, {P{\'e}rez-Fournon}, {Pohlen}, {Popesso},
  {Pozzi}, {Rigopoulou}, {Riguccini}, {Rosario}, {Roseboom}, {Rowan-Robinson},
  {Salvato}, {Schulz}, {Scott}, {Seymour}, {Shupe}, {Smith}, {Valtchanov},
  {Wang}, {Xu}, {Zemcov}, \& {Wuyts}}]{2013MNRAS.431.2317S}
{Symeonidis}, M., {Vaccari}, M., {Berta}, S., {et~al.} 2013, \mnras, 431, 2317

\bibitem[{{Vega} {et~al.}(2008){Vega}, {Clemens}, {Bressan}, {Granato},
  {Silva}, \& {Panuzzo}}]{2008A&A...484..631V}
{Vega}, O., {Clemens}, M.~S., {Bressan}, A., {et~al.} 2008, \aap, 484, 631

\bibitem[{{Vieira} {et~al.}(2013){Vieira}, {Marrone}, {Chapman}, {De Breuck},
  {Hezaveh}, {Wei{$\beta$}}, {Aguirre}, {Aird}, {Aravena}, {Ashby}, {Bayliss},
  {Benson}, {Biggs}, {Bleem}, {Bock}, {Bothwell}, {Bradford}, {Brodwin},
  {Carlstrom}, {Chang}, {Crawford}, {Crites}, {de Haan}, {Dobbs}, {Fomalont},
  {Fassnacht}, {George}, {Gladders}, {Gonzalez}, {Greve}, {Gullberg},
  {Halverson}, {High}, {Holder}, {Holzapfel}, {Hoover}, {Hrubes}, {Hunter},
  {Keisler}, {Lee}, {Leitch}, {Lueker}, {Luong-van}, {Malkan}, {McIntyre},
  {McMahon}, {Mehl}, {Menten}, {Meyer}, {Mocanu}, {Murphy}, {Natoli}, {Padin},
  {Plagge}, {Reichardt}, {Rest}, {Ruel}, {Ruhl}, {Sharon}, {Schaffer}, {Shaw},
  {Shirokoff}, {Spilker}, {Stalder}, {Staniszewski}, {Stark}, {Story},
  {Vanderlinde}, {Welikala}, \& {Williamson}}]{2013Natur.495..344V}
{Vieira}, J.~D., {Marrone}, D.~P., {Chapman}, S.~C., {et~al.} 2013, \nat, 495,
  344

\bibitem[{{Walter} {et~al.}(2012){Walter}, {Decarli}, {Carilli}, {Bertoldi},
  {Cox}, {da Cunha}, {Daddi}, {Dickinson}, {Downes}, {Elbaz}, {Ellis}, {Hodge},
  {Neri}, {Riechers}, {Weiss}, {Bell}, {Dannerbauer}, {Krips}, {Krumholz},
  {Lentati}, {Maiolino}, {Menten}, {Rix}, {Robertson}, {Spinrad}, {Stark}, \&
  {Stern}}]{2012Natur.486..233W}
{Walter}, F., {Decarli}, R., {Carilli}, C., {et~al.} 2012, \nat, 486, 233

\bibitem[{{Wei{\ss}} {et~al.}(2013){Wei{\ss}}, {De Breuck}, {Marrone},
  {Vieira}, {Aguirre}, {Aird}, {Aravena}, {Ashby}, {Bayliss}, {Benson},
  {B{\'e}thermin}, {Biggs}, {Bleem}, {Bock}, {Bothwell}, {Bradford}, {Brodwin},
  {Carlstrom}, {Chang}, {Chapman}, {Crawford}, {Crites}, {de Haan}, {Dobbs},
  {Downes}, {Fassnacht}, {George}, {Gladders}, {Gonzalez}, {Greve},
  {Halverson}, {Hezaveh}, {High}, {Holder}, {Holzapfel}, {Hoover}, {Hrubes},
  {Husband}, {Keisler}, {Lee}, {Leitch}, {Lueker}, {Luong-Van}, {Malkan},
  {McIntyre}, {McMahon}, {Mehl}, {Menten}, {Meyer}, {Murphy}, {Padin},
  {Plagge}, {Reichardt}, {Rest}, {Rosenman}, {Ruel}, {Ruhl}, {Schaffer},
  {Shirokoff}, {Spilker}, {Stalder}, {Staniszewski}, {Stark}, {Story},
  {Vanderlinde}, {Welikala}, \& {Williamson}}]{2013ApJ...767...88W}
{Wei{\ss}}, A., {De Breuck}, C., {Marrone}, D.~P., {et~al.} 2013, \apj, 767, 88

\end{thebibliography}

\end{document}